\documentclass[amsmath,amssymb,showkeys,superscriptaddress,floatfix,aps,prd,10pt,twocolumn]{revtex4}
\usepackage{graphicx,epstopdf}
\usepackage{rotating}
\usepackage{tabularx}
\usepackage[caption=false]{subfig}
\usepackage{multirow}
\usepackage{appendix}
\usepackage{xcolor}
\usepackage[colorlinks=true, urlcolor=blue, linkcolor=blue, citecolor=green]{hyperref}

\def\pslash{p\!\!\!\slash }
\def\qslash{q\!\!\!\slash }
\def\xslash{x\!\!\!\slash }
\def\eslash{\varepsilon\!\!\!\slash }

\def\vel{\left|}
\def\ver{\right|}

\begin{document}

\title{Magnetic dipole moments of the spin-$\frac{3}{2}$ doubly heavy baryons}

\author{Ula\c{s} \"{O}zdem}%
\email[]{ulasozdem@aydin.edu.tr}
\affiliation{
 Health Services Vocational School of Higher Education, Istanbul Aydin University, Sefakoy-Kucukcekmece, 34295 Istanbul, Turkey
}

\begin{abstract}
The magnetic dipole moments of the spin-$\frac{3}{2}$  doubly charmed, bottom and charmed-bottom baryons  are obtained by means of the light-cone QCD sum rule. 
The magnetic dipole moments of these baryons encode essential knowledge of their inner structure and shape deformations. The numerical results are given as,
$\mu_{\Xi_{cc}^{*++}} = 2.94 \pm 0.95$, $\mu_{\Xi_{cc}^{*+}} = - 0.67 \pm 0.11$, $\mu_{\Omega_{cc}^{*+}} =- 0.52 \pm 0.07$, $\mu_{\Xi_{bb}^{*0}} = 2.30 \pm 0.55$, 
$\mu_{\Xi_{bb}^{*-}} = -1.39 \pm 0.32$, $\mu_{\Omega_{bb}^{*-}} = -1.56 \pm 0.33$, $\mu_{\Xi_{bc}^{*+}} = 2.63 \pm 0.82$, $\mu_{\Xi_{bc}^{*0}} = - 0.96 \pm 0.32$  
and $\mu_{\Omega_{bc}^{*+}} =- 1.11 \pm 0.33$, respectively. 
\end{abstract}

\maketitle

\section{Motivation}
The doubly heavy baryons (DHBs) presumably contain two heavy quark and one light quark. 
One of them was firstly announced by the SELEX
Collaboration in the decay mode $\Xi_{cc}^+ \rightarrow \Lambda_c^+ K^- \pi^+$
with the mass $M_{\Xi^+_{cc}}=3519 \pm 1$~MeV~\cite{Mattson:2002vu}. 
However, neither Belle~\cite{Chistov:2006zj}, nor FOCUS~\cite{Ratti:2003ez},
 nor BABAR~\cite{Aubert:2006qw} could confirm  the DHBs 
 in $e^-\,e^+$ annihilations.
 It is worth pointing out that the analysis of the SELEX experiment with other experimental 
 groups is achieved through different production mechanisms. 
 Therefore, the results of the SELEX Collaboration cannot be ruled out.
In 2017, LHCb Collaboration observed another doubly heavy baryon $\Xi_{cc}^{++}$ in the mass spectrum of 
$\Lambda_c^+\,K^-\pi^+\,\pi^+ $ with the mass $M_{\Xi_{cc}^{++}}=3621.40\pm 0.72\pm0.27\pm0.14$~MeV~\cite{Aaij:2017ueg}.
Recently, the LHCb Collaboration reconstructed their analysis via decay modes $ \Xi_{cc}^{++} \rightarrow \Lambda_c^+\,K^-\pi^+\,\pi^+ $ and $\Xi_{cc}^{++} \rightarrow \Xi_{c}^{+} \,\pi^+$ and they reported their mass value as 
$M_{\Xi_{cc}^{++}}=3621.55\pm 0.23\pm0.30$~MeV~\cite{Aaij:2019uaz}.
The investigation for the DHBs may provide with valuable knowledge for comprehension of the nonperturbative
QCD effects. One of the several point of views which makes the physics of DHBs charming is that the binding of two charm quarks and
a light quark provides a unique perspective for dynamics of confinement. 
%
The research of the properties of DHBs is one of the active and interesting branches of particle physics.
%
%
Therefore, the weak~\cite{Albertus:2006ya,Li:2017ndo,Wang:2017mqp,Wang:2017azm,Shi:2017dto,Shi:2019hbf,Shi:2019fph}, strong~\cite{Hu:2005gf,Xiao:2017udy}  and radiative decays~\cite{Li:2017pxa,Yu:2017zst,Lu:2017meb,Cui:2017udv}, the magnetic dipole moments (MDMs)~\cite{Can:2013zpa,Branz:2010pq,Bose:1980vy,Patel:2008xs,
SilvestreBrac:1996bg,Patel:2007gx,Gadaria:2016omw,JuliaDiaz:2004vh,Faessler:2006ft,Can:2013tna,Li:2017cfz,Bernotas:2012nz,Lichtenberg:1976fi,Oh:1991ws,Simonis:2018rld,Liu:2018euh,Blin:2018pmj,Meng:2017dni,Dhir:2009ax, Ozdem:2018uue} and masses~\cite{Bagan:1992za,Roncaglia:1995az,Ebert:1996ec,Tong:1999qs,Itoh:2000um,Gershtein:2000nx,
Kiselev:2001fw,Kiselev:2002iy,Narodetskii:2001bq,Lewis:2001iz,Ebert:2002ig,Mathur:2002ce,Flynn:2003vz,Vijande:2004at,
Chiu:2005zc,Migura:2006ep,Albertus:2006ya,Martynenko:2007je,Tang:2011fv,Liu:2007fg,Roberts:2007ni,
Valcarce:2008dr,Liu:2009jc,Alexandrou:2012xk,Aliev:2012ru,Aliev:2012iv,Namekawa:2013vu,Karliner:2014gca,Sun:2014aya,
Chen:2015kpa,Sun:2016wzh,Shah:2016vmd,Kiselev:2017eic,Chen:2017sbg,Hu:2005gf,Meng:2017fwb,
Narison:2010py,Zhang:2008rt,Guo:2017vcf,Lu:2017meb,Xiao:2017udy,Weng:2018mmf,Can:2013zpa,Branz:2010pq,Bose:1980vy,Patel:2008xs,SilvestreBrac:1996bg,Patel:2007gx,Gadaria:2016omw}  of the DHBs have been examined broadly in literature by the help of the quark models, potential models, lattice QCD,  Feynman-Hellmann theorem,  extended chromomagnetic model, 
chiral perturbation theory, heavy quark effective theory,  QCD sum rules,  perturbative QCD, Faddeev approach,  SU(3) flavor symmetry, nonperturbative string approach, local diquark approach, light-cone QCD sum rules, light front approach and extended on-mass-shell renormalization scheme. 

Electromagnetic properties one of the major and meaningful parameters of the DHBs.
As the electromagnetic properties symbolise necessary viewpoints of the intrinsic properties of hadrons, it is quite
significant to analyze the baryon electromagnetic form factors, particularly the MDMs.
The magnitude and sign of the MDM ensure crucial 
data on size, structure and shape deformations of baryons. 
Apparently, determining the MDM is an important step in our comprehension of the baryon features with regards to quark-gluon degrees of freedom.
In this study, we are going to concentrate on the DHBs (from now on we will represent
these particles as $B^{*}_{QQ}$ ) with spin-parity $J^{P} = \frac{3}{2}^{+}$, and calculate their 
MDMs by the help of the light-cone QCD sum rule (LCSR) approach, 
which is one of the powerful nonperturbative methods in hadron physics providing us to calculate properties of the particles and processes. 
In LCSR, the hadronic properties are
expressed with regards to the vacuum condensates and the light-cone distribution amplitudes  of the on-shell particles \cite{Chernyak:1990ag, Braun:1988qv, Balitsky:1989ry}. 
Since the MDMs are quantities in terms of the properties of the vacuum and distribution amplitudes of the particles, any uncertainties
in these parameters are reflected to the uncertainties of the predictions of the MDMs. 
%
The first extraction of the MDMs of the spin-$\frac{3}{2}$ doubly heavy
baryons were obtained by Lichtenberg~\cite{Lichtenberg:1976fi} by means of the naive quark model. 
Then, many scientists have used different theoretical models to compute the spin-$\frac{3}{2}$ doubly heavy baryon MDMs~\cite{Albertus:2006ya,Shah:2016vmd, Shah:2017liu,Meng:2017dni, Dhir:2009ax,Dhir:2013nka,Bernotas:2012nz,Simonis:2018rld,Gadaria:2016omw}.  
In Ref.~\cite{Albertus:2006ya}, they have been calculated the static properties and semileptonic decays of the DHBs with the help of nonrelativistic quark model. 
To examination the dependence of their results on inter-quark interaction they use various quark potential that contain hyperfine terms coming from one gluon exchange and Coulomb term as well.
In Refs.~\cite{Shah:2016vmd, Shah:2017liu}, the masses of the ground, orbital and radial states of the DHBs are evaluated 
in the framework of hypercentral constituent quark model with Coulomb plus linear potential.
The MDMs of the ground state DHBs are also extracted.
In Ref.~\cite{Meng:2017dni}, the MDMs of the spin-$\frac{3}{2}$ doubly charmed baryons are investigated up to NNLO by means of the heavy baryon chiral perturbation theory.
As a by product, they obtained the numerical values of the doubly bottom and charmed-bottom baryons up to NLO.
In Refs.~\cite{Dhir:2009ax,Dhir:2013nka}, they are calculated the MDMs of singly and DHBs with spin-$\frac{3}{2}$ and spin-$\frac{1}{2}$ 
in the framework of effective quark mass and screened charge of quark.
In Ref.~\cite{Bernotas:2012nz}, the MDMs of the DHBs are investigated in the MIT bag model with center of mass motion corrections. 
In Ref.~\cite{Simonis:2018rld}, the decay widths,  MDMs and M1 transitions of the all ground states heavy baryons are evaluated by means of the  extended MIT bag model.
In Ref.~\cite{Gadaria:2016omw}, they are utilized the extended relativistic harmonic confinement model to acquire masses and MDMs of the heavy flavored ground state baryons.

The plan of the manuscript is as follows.
In section \ref{secII}, the details of the MDMs computations
for the DHBs with spin-$\frac{3}{2}$ are presented. 
Numerical analysis of the LCSR for the MDMs are given in section \ref{secIII}.
Section \ref{secIV} is reserved for discussion and concluding remarks.

\section{Formalism} \label{secII}
To obtain the MDM of the DHBs by using the
LCSR approach, we begin with the subsequent correlation function,

\begin{eqnarray}
 \label{edmn00}
\Pi _{\mu \nu \alpha }(p,q)&=&i^2\int d^{4}x\,\int d^{4}y\,e^{ip\cdot x+iq \cdot y}\nonumber\\
&&\times \langle 0|\mathcal{T}\{J_{\mu}^{B^{*}_{QQ}}(x) J_{\alpha}(y)
J_{\nu }^{B^{*\dagger}_{QQ}}(0)\}|0\rangle. 
\end{eqnarray}%
 Here, $J_{\mu(\nu)}$ is the interpolating current of the $B^{*}_{QQ}$ baryons and 
 the electromagnetic current $J_\alpha$ is given as,
\begin{equation}
 J_\alpha =\sum_{q= u,d,s,c,b} e_q \bar q \gamma_\alpha q,
\end{equation}
where $e_q$ is the electric charge of the corresponding quark.


From a technical point of view, 
the Eq. (\ref{edmn00}) can be 
reconsidered in a more appropriate form by the help of external background electromagnetic (EBGEM) field,

\begin{align}
 \label{edmn01}
\Pi _{\mu \nu }(p,q)=i\int d^{4}x\,e^{ip\cdot x}
\langle 0|\mathcal{T}\{J_{\mu}^{B^{*}_{QQ}}(x)
J_{\nu }^{B^{*\dagger}_{QQ}}(0)\}|0\rangle_{F}, 
\end{align}%
where  F is the EBGEM field and 
$F_{\alpha\beta}= i (\varepsilon_\alpha q_\beta-\varepsilon_\beta q_\alpha)$ with
 $\varepsilon_\beta$ and $q_\alpha$  being the polarization and four-momentum 
of the EBGEM field, respectively.
Since the EBGEM field can be made arbitrarily small,
the correlation function in Eq. (\ref{edmn01}) can be acquired by expanding in powers of the EBGEM field, 
 \begin{equation}
\Pi _{\mu \nu }(p,q) = \Pi _{\mu \nu }^{(0)}(p,q) + \Pi _{\mu \nu }^{(1)}(p,q)+.... ,
\end{equation}
and keeping only terms $\Pi _{\mu \nu }^{(1)}(p,q)$, which corresponds 
to the single photon emission~\cite{Ball:2002ps}.
The primary advantage of using the EBGEM field approach
relies on the fact that it separates the soft and hard photon emissions in an explicitly gauge 
invariant way~\cite{Ball:2002ps}.
The $\Pi _{\mu \nu }^{(0)}(p,q)$ is the correlation function in the 
lack of the EBGEM field, and leads to the two point
sum rules of the hadrons, which is not relevant for our case.

After these general comments, we can now move on deriving the LCSR
for the MDM of the DHBs.
The correlation function given in Eq. (\ref{edmn01}) can be calculated 
with regards to hadronic properties, known as hadronic representation. 
In addition to this it can be obtained with regards to the quark-gluon properties 
in the deep Euclidean region, known as QCD representation.
By matching the results of these representations using the dispersion relation and quark–hadron duality ansatz, one can acquire the corresponding sum rules.

We begin to evaluate the correlation function with respect to hadronic degrees of freedom comprising the physical properties of the particles under investigation. 
For that purpose, we embed a complete set of $B^{*}_{QQ}$ baryons into the
correlation function. So, we get
\begin{align}\label{edmn16}
\Pi^{Had}_{\mu\nu}(p,q)&=\frac{\langle0\mid J_{\mu}^{B^*_{QQ}}\mid
{B^{*}_{QQ}}(p)\rangle}{[p^{2}-m_{{B^{*}_{QQ}}}^{2}]}\nonumber\\
&\langle {B^{*}_{QQ}}(p)\mid
{B^{*}_{QQ}}(p+q)\rangle_F \nonumber\\
&\frac{\langle {B^{*}_{QQ}}(p+q)\mid
\bar{J}_{\nu}^{B^*_{QQ}}\mid 0\rangle}{[(p+q)^{2}-m_{{B^{*}_{QQ}}}^{2}]}+...,
\end{align}
where the dots stand for contributions of higher states and the continuum.
The matrix elements in Eq. (\ref{edmn16}) are defined as~\cite{Pascalutsa:2006up,Ramalho:2009vc},
\begin{align}\label{edmn17}
\langle0\mid J_{\mu}(0)\mid {B^{*}_{QQ}}(p,s)\rangle&=\lambda_{{B^{*}_{QQ}}}u_{\mu}(p,s),\\
\langle {B^{*}_{QQ}}(p)\mid {B^{*}_{QQ}}(p+q)\rangle_F &=-e\,\bar
u_{\mu}(p)\Bigg\{F_{1}(q^2)g_{\mu\nu}\eslash\nonumber\\
&-\frac{1}{2m_{{B^{*}_{QQ}}}}
\Big[F_{2}(q^2)g_{\mu\nu}+F_{4}(q^2)\nonumber\\
& \times \frac{q_{\mu}q_{\nu}}{(2m_{{B^{*}_{QQ}}})^2}\Big]\eslash\qslash+\frac{F_{3}(q^2)}{(2m_{{B^{*}_{QQ}}})^2}
\nonumber\\
& \times q_{\mu}q_{\nu}\eslash\Bigg\} u_{\nu}(p+q),
\end{align}
where $\lambda_{{B^{*}_{QQ}}}$ is the  residue of ${B^{*}_{QQ}}$ baryon  and $u_{\mu}(p,s)$ is the
Rarita-Schwinger spinor.  Summation over spins of ${B^{*}_{QQ}}$ baryon is carried out as:

\begin{eqnarray}\label{raritabela}
\sum_{s}u_{\mu}(p,s)\bar u_{\nu}(p,s)&=&-\Big(\pslash+m_{{B^{*}_{QQ}}}\Big)\Big[g_{\mu\nu} -\frac{1}{3}\gamma_{\mu}\gamma_{\nu}\nonumber\\
&&-\frac{2\,p_{\mu}p_{\nu}}
{3\,m^{2}_{{B^{*}_{QQ}}}}+\frac{p_{\mu}\gamma_{\nu}-p_{\nu}\gamma_{\mu}}{3\,m_{{B^{*}_{QQ}}}}\Big].
\end{eqnarray}
Substituting Eqs. (\ref{edmn16})-(\ref{raritabela}) into Eq. (\ref{edmn01}) for hadronic side we obtain 
 %
\begin{align}\label{fizson}
 \Pi^{Had}_{\mu\nu}(p,q)&=-\frac{\lambda_{_{{B^{*}_{QQ}}}}^{2}\,\Big(\pslash+m_{{B^{*}_{QQ}}}\Big)}{[(p+q)^{2}-m_{_{{B^{*}_{QQ}}}}^{2}][p^{2}-m_{_{{B^{*}_{QQ}}}}^{2}]}
 \nonumber\\
 &
 \Big[g_{\mu\nu}
-\frac{1}{3}\gamma_{\mu}\gamma_{\nu}-\frac{2\,p_{\mu}p_{\nu}}
{3\,m^{2}_{{B^{*}_{QQ}}}}+\frac{p_{\mu}\gamma_{\nu}-p_{\nu}\gamma_{\mu}}{3\,m_{{B^{*}_{QQ}}}}\Big]\nonumber\\
&\times \Bigg\{F_{1}(q^2)g_{\mu\nu}\eslash -
\frac{1}{2m_{{B^{*}_{QQ}}}}
\Big[F_{2}(q^2)g_{\mu\nu}\nonumber\\
&+F_{4}(q^2) \frac{q_{\mu}q_{\nu}}{(2m_{{B^{*}_{QQ}}})^2}\Big]\eslash\qslash+\frac{F_{3}(q^2)}{(2m_{{B^{*}_{QQ}}})^2}
 q_{\mu}q_{\nu}\eslash\Bigg\}.
\end{align}

As a principle, we can achieve the last form of the hadronic representation of the correlator using the Eq. (\ref{fizson}), however we face with two problems.
 One of them is related to the fact that not all Lorentz structures appearing in Eq. (\ref{fizson}) are independent. 
The second problem is 
the correlator can also get contributions from spin-1/2 baryons, which should be removed. 
To eliminate the spin one half contributions and acquire just independent structures in the
correlator, we perform the  ordering for Dirac matrices as $\gamma_{\mu}\pslash\eslash\qslash\gamma_{\nu}$ and remove terms 
with $\gamma_\mu$ at the beginning, $\gamma_\nu$ at the end and all those proportional to $p_\mu$ and 
$p_\nu$~\cite{Belyaev:1982cd}. As a result, 
for hadronic side we get,
\begin{align}\label{final phenpart}
\Pi^{Had}_{\mu\nu}(p,q)=&-\frac{\lambda_{_{{B^{*}_{QQ}}}}^{2}}{[(p+q)^{2}-m_{_{{B^{*}_{QQ}}}}^{2}][p^{2}-m_{_{{B^{*}_{QQ}}}}^{2}]}\nonumber\\
&\times
\Bigg[  -g_{\mu\nu}\pslash\eslash\qslash \,F_{1}(q^2) 
+ m_{{B^{*}_{QQ}}}g_{\mu\nu}\eslash\qslash\,F_{2}(q^2)\nonumber\\
&+~
\mbox{other independent structures} \Bigg].
\end{align}

The MDM form factor, $G_{M}(q^2)$,  is
defined with respect to the form factors $F_{i}(q^2)$ in the following
manner~\cite{Pascalutsa:2006up,Ramalho:2009vc}:
\begin{eqnarray}
G_{M}(q^2) &=& \big[ F_1(q^2) + F_2(q^2)\big] ( 1+ \frac{4}{5}
\tau ) -\frac{2}{5} \big[ F_3(q^2)\nonumber\\
&&+F_4(q^2)\big] \tau ( 1 + \tau ), 
\end{eqnarray}
  where $\tau
= -\frac{q^2}{4m^2_{{B^{*}_{QQ}}}}$. At $q^2=0$, the magnetic dipole form factors
are obtained with respect tothe functions $F_i(0)$ as:
\begin{eqnarray}\label{mqo1}
G_{M}(0)&=&F_{1}(0)+F_{2}(0).
\end{eqnarray}
The  MDM ($\mu_{{B^{*}_{QQ}}}$), is defined in the following way:
 \begin{eqnarray}\label{mqo2}
\mu_{{B^{*}_{QQ}}}&=&\frac{e}{2m_{{B^{*}_{QQ}}}}G_{M}(0).
\end{eqnarray}

In this study we achieve QCD sum rules for the form factors $ F_i(q^2) $ at first, after that in numerical calculations we will make use of the above equations to obtain the values of the MDMs using the QCD sum rules for the form factors. 
The final form of the hadronic representation with respect to the chosen structures in momentum space is:
\begin{eqnarray}
\Pi^{Had}_{\mu\nu}(p,q)&=&\Pi_1^{Had}g_{\mu\nu}\pslash\eslash\qslash \,
+\Pi_2^{Had}g_{\mu\nu}\eslash\qslash\,+
...,
\end{eqnarray}
where $ \Pi_i^{Had} $ are functions of the form factors $ F_i(q^2) $ and other hadronic parameters; and $ ... $ represents other independent structures.

To obtain the expression of the correlation function with respect to the quark-gluon parameters, the explicit form
for the interpolating current of the $B^*_{QQ}$ baryons needs to be chosen.
In this work, we consider the $B^*_{QQ}$ baryons with the quantum numbers $J^P =\frac{3}{2}^+$. 
The interpolating current is given as
~\cite{Aliev:2012iv},

\begin{align}
 \label{_curr}
J_{\mu}^{B^*_{QQ}}(x)&= {\frac{1}{\sqrt{3}}} \epsilon^{abc} \Big\{
(q^{aT} C \gamma_\mu Q^b) Q^{\prime c} +(q^{aT} C \gamma_\mu Q^{\prime b}) Q^c\nonumber\\
&
+(Q^{aT} C \gamma_\mu Q^{\prime b}) q^c \Big\}~,
\end{align}
where  $q$ is the light; and $Q$ and $Q^\prime$
are the two  heavy quarks, respectively. We give the quark content of the spin-3/2 DHBs in Table \ref{Table_curr}.
\begin{table}[htp]
\caption{The quark content of the spin-3/2 DHBs.}
\renewcommand{\arraystretch}{1.3}
\addtolength{\arraycolsep}{-0.5pt}
\small
$$
\begin{array}{l|c|c|cc}
\hline \hline   
  \mbox{Baryon}           & \mbox{ $q$} & \mbox{ $Q$}  & \mbox{ $Q^\prime$}\\  \hline\hline
 \Xi_{Q Q}^\ast           & u~\mbox{or}~d          & b~\mbox{or}~c           & b~\mbox{or}~c               \\
 \Xi_{Q Q^\prime}^\ast    & u~\mbox{or}~d          & b                       & c                            \\
 \Omega_{Q Q}^\ast        & s                      & b~\mbox{or}~c           &b~\mbox{or}~c                \\
  \Omega_{Q Q^\prime}^\ast & s                      & b                       &            c                \\
\hline \hline
\end{array}
$$
\renewcommand{\arraystretch}{1}
\addtolength{\arraycolsep}{-0.5pt}
\label{Table_curr}
\end{table}

After contracting pairs of quark fields and using the Wick's theorem, the correlation function becomes:

\begin{widetext}
\begin{eqnarray}
\label{edmn18}
\Pi^{QCD}_{\mu\nu}(p)&=&-\frac{i}{3}\,\varepsilon^{abc}\varepsilon^{a^{\prime}b^{\prime}c^{\prime}}
\int d^4x e^{ip\cdot x} \langle 0| \Bigg\{
 S_{Q^\prime}^{c c^\prime}  \mbox{Tr}\Big[S_{Q}^{b a^\prime} \gamma_\nu
\widetilde{S}_{q}^{a b^\prime} \gamma_\mu \Big] +
S_{q}^{c c^\prime}  \mbox{Tr}\Big[S_{Q^\prime}^{b a^\prime} \gamma_\nu
\widetilde{S}_{Q}^{a b^\prime} \gamma_\mu \Big] \nonumber\\
&&+
S_{Q}^{c c^\prime}  \mbox{Tr}\Big[S_{q}^{b a^\prime} \gamma_\nu
\widetilde{S}_{Q^\prime}^{a b^\prime} \gamma_\mu \Big]
+
S_Q^{c b^\prime} \gamma_\nu \widetilde{S}_{Q^\prime}^{a a^\prime} \gamma_\mu
S_q^{b c^\prime} + S_Q^{c a^\prime} \gamma_\nu \widetilde{S}_{q}^{b b^\prime}
\gamma_\mu S_{Q^\prime}^{a c^\prime}
\nonumber\\
&& 
+
S_{Q^\prime}^{c a^\prime} \gamma_\nu \widetilde{S}_{Q}^{b b^\prime} \gamma_\mu
S_{q}^{a c^\prime} +S_{Q^\prime}^{c b^\prime} \gamma_\nu \widetilde{S}_{q}^{a a^\prime}
\gamma_\mu S_{Q}^{b c^\prime}
 + S_q^{c a^\prime} \gamma_\nu
\widetilde{S}_{Q^\prime}^{b b^\prime} \gamma_\mu S_Q^{a c^\prime}\nonumber \\
&&  +
S_q^{c b^\prime} \gamma_\nu
\widetilde{S}_{Q}^{a a^\prime} \gamma_\mu S_{Q^\prime}^{b c^\prime} 
 \Bigg\}|0 \rangle_F ,
\end{eqnarray}
\end{widetext}
where $\tilde S_{Q(q)}^{ij}(x) = CS_{Q(q)}^{{ij}^T}(x)C$ and, $S_q^{ij}(x)$ and
$S_Q^{ij}(x)$ are the light and heavy quark propagators, respectively.
The light and heavy quark propagators are given as~\cite{Yang:1993bp, Belyaev:1985wza},
\begin{align}
\label{edmn09}
S_{q}(x)&=S_q^{free}
- \frac{ \bar qq }{12} \Big(1-i\frac{m_{q} \xslash}{4}   \Big)
- \frac{\bar q \sigma.G q }{192}x^2  \Big(1-i\frac{m_{q} \xslash}{6}   \Big)\nonumber\\
&
-\frac {i g_s }{32 \pi^2 x^2} ~G^{\mu \nu} (x) \bigg[\rlap/{x}
\sigma_{\mu \nu} +  \sigma_{\mu \nu} \rlap/{x}
 \bigg],\nonumber\\
S_{Q}(x)&=S_Q^{free}
-\frac{g_{s}m_{Q}}{16\pi ^{2}} \int_{0}^{1}dv~G^{\mu \nu }(vx)\bigg[ \big(\sigma _{\mu \nu }{\xslash}
  +{\xslash}\sigma _{\mu \nu }\big)\nonumber\\
  &\times \frac{K_{1}( m_{Q}\sqrt{-x^{2}}) }{\sqrt{-x^{2}}}
+2\sigma ^{\mu \nu }K_{0}( m_{Q}\sqrt{-x^{2}})\bigg],
\end{align}%
where 
\begin{align}
 &S_q^{free} =\frac{1}{2 \pi^2 x^2}\Big( i \frac{{\xslash}}{x^{2}}-\frac{m_{q}}{2 } \Big),\nonumber\\
&S_Q^{free} = \frac{m_{Q}^{2}}{4 \pi^{2}} \bigg[ \frac{K_{1}(m_{Q}\sqrt{-x^{2}}) }{\sqrt{-x^{2}}}
+i\frac{{\xslash}~K_{2}( m_{Q}\sqrt{-x^{2}})}
{(\sqrt{-x^{2}})^{2}}\bigg],
\end{align}
with $G^{\mu \nu }$ is the gluon field strength tensor, $K_{i}$ are second kind of the Bessel functions, $m_q$ and $m_Q$ are the light
and heavy quark mass respectively.

\begin{figure}[t]
\centering
\subfloat[]{\includegraphics[width=0.32\textwidth]{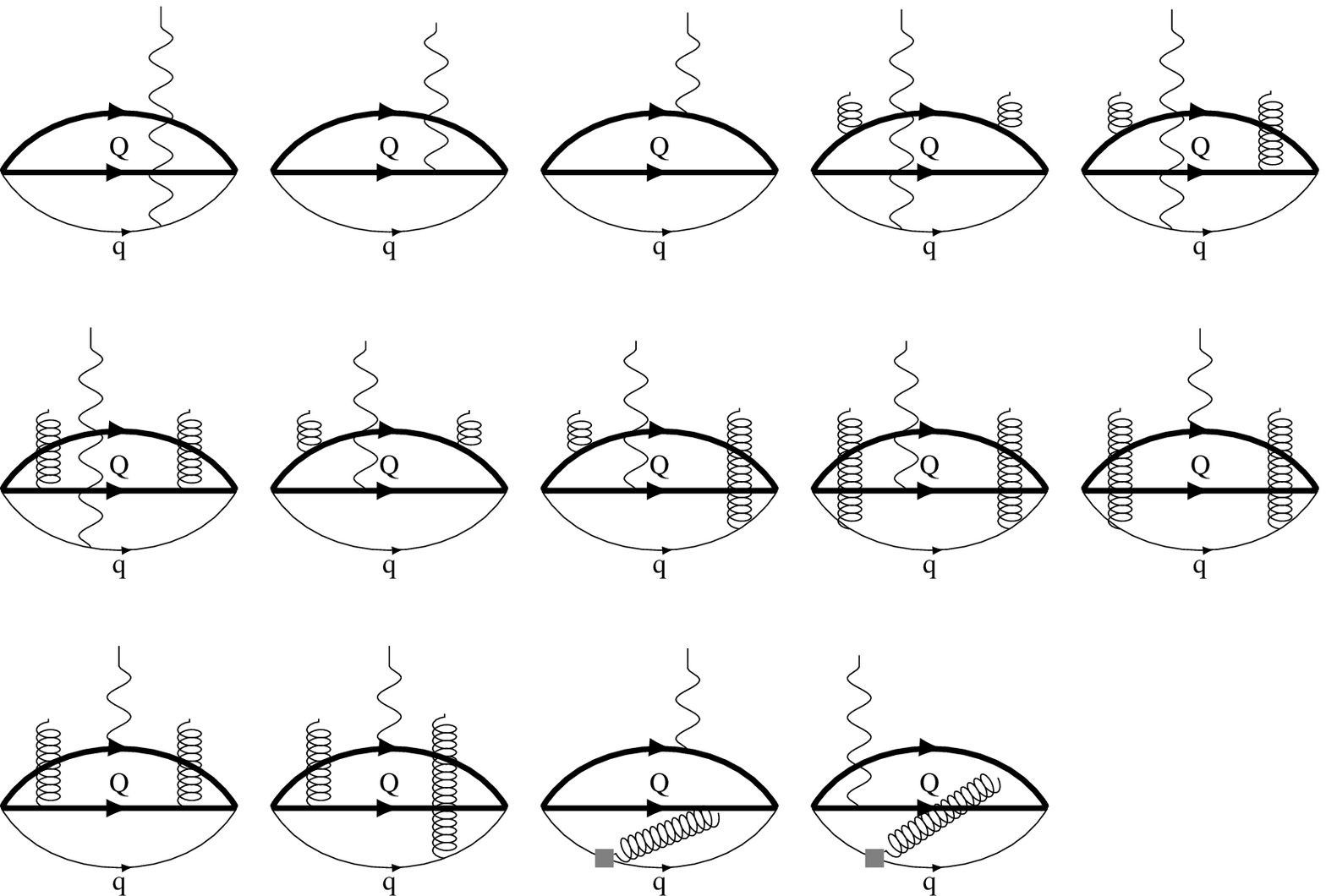}}\\
\subfloat[]{\includegraphics[width=0.32\textwidth]{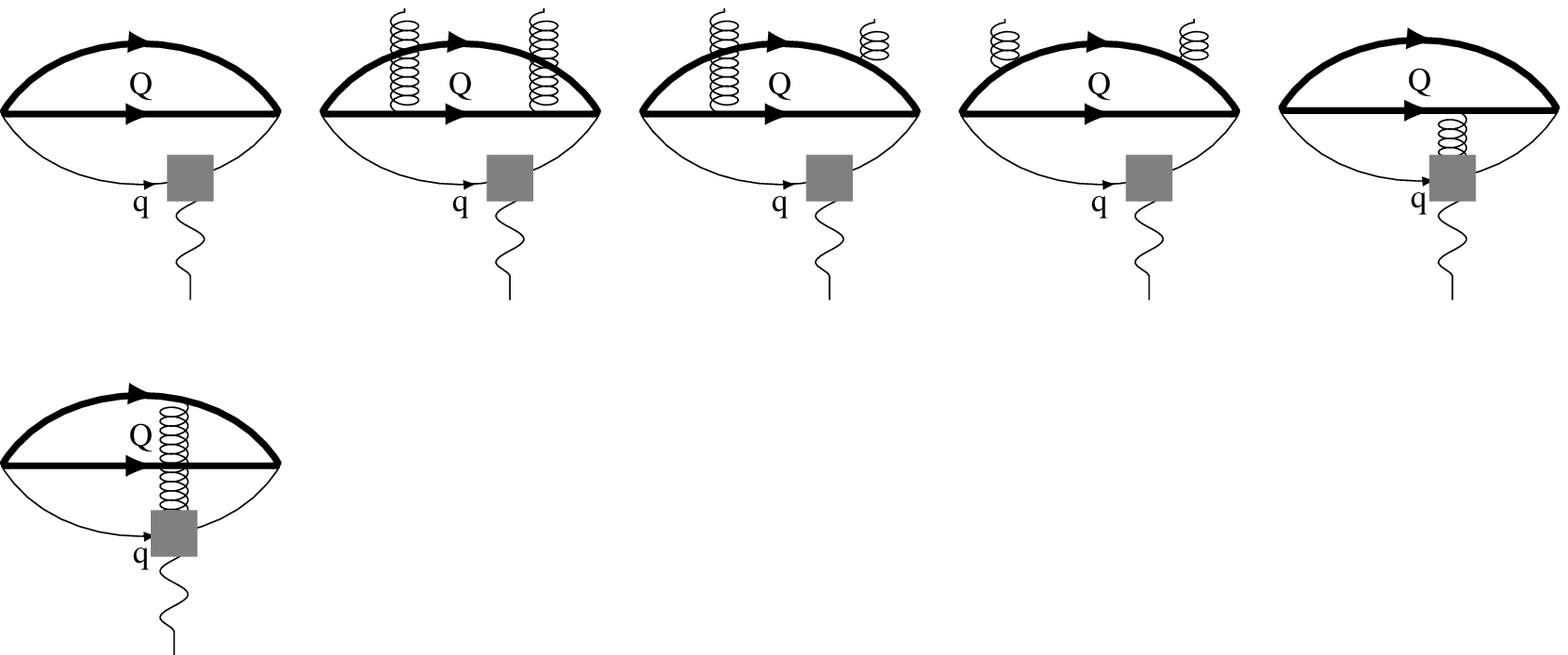}}
 \caption{ Feynman diagrams for the MDMs of the spin-3/2 DHBs. The thick, thin, wavy and
curly lines denote the heavy quark, light quark, photon and gluon propagators, respectively. 
 Diagrams (a) corresponding to the perturbative photon vertex and,  
 diagrams (b) denote the contributions coming from the distribution amplitudes of the photon.}
  \end{figure}

  The correlator in Eq. (\ref{edmn18}) contains various contributions: the photon can be emitted both perturbatively or nonperturbatively.
  When the photon is emitted perturbatively, one of the propagators in Eq. (\ref{edmn18}) is substituted by
 \begin{eqnarray}
\label{edmn12}
S^{free} (x) \rightarrow \int d^4y\, S^{free} (x-y)\,\rlap/{\!A}(y)\, S^{free} (y)\,,
\end{eqnarray}
 and the surviving two propagators are substituted with the
full quark propagators including the free (perturbative) part as well as the
interacting parts (with gluon or QCD vacuum) as nonperturbative contributions.
The total perturbative photon emission is achieved by carrying out the
substitution mentioned above for the perturbatively interacting quark propagator with the photon and
employing the substitution of the surviving propagators by their free parts.

In case of nonperturbative photon emission, the light quark propagator in Eq. (\ref{edmn18}) is substituted by
\begin{align}
\label{edmn13}
S_{\alpha\beta}^{ab} \rightarrow -\frac{1}{4} (\bar{q}^a \Gamma_i q^b)(\Gamma_i)_{\alpha\beta},
\end{align}
where $\Gamma_i$ represent the full set of Dirac matrices. 
Under this approach, two surviving quark propagators are taken as the full propagators comprising perturbative as well as nonperturbative contributions.
 Once Eq. (\ref{edmn13}) is inserted into Eq. (\ref{edmn18}), there seem matrix
elements such as $\langle \gamma(q)\vel \bar{q}(x) \Gamma_i q(0) \ver 0\rangle$
and $\langle \gamma(q)\vel \bar{q}(x) \Gamma_i G_{\alpha\beta}q(0) \ver 0\rangle$,
representing the nonperturbative contributions. 
Furthermore, nonlocal operators such as $\bar{q}q\bar{q}q$  and 
$\bar{q} G^2 q$ are anticipated to appear.
%
In this study, we take into account operators includes only one gluon field and three particle nonlocal operators and to disregard terms with four quarks $\bar{q}q\bar{q}q$, and two gluons $\bar{q} G^2 q$.
In order to calculate the nonperturbative contributions,
we need the matrix elements of the nonlocal operators between the photon states  and the vacuum and
these matrix elements are described with respect to the photon distribution amplitudes with definite
twists. Up to twist-4 the explicit expressions of the photon distribution amplitudes are given in~\cite{Ball:2002ps}.
%
%
Using these expressions for the propagators and distribution amplitudes for the photon, the correlation functions from the QCD side can be computed.

The QCD and hadronic representations of the correlation function are then matched using dispersion relation.
The next step in deriving the sum rules for the MDMs of the spin-$\frac{3}{2}$ DHBs is applying double Borel transformations (${\cal B}  $)
over the $p^2$ and $(p+q)^2$ on the both representations of the correlation function 
so as to suppress the contributions of higher states and continuum.
As a result, we obtain

\begin{eqnarray}
{\cal B}\Pi^{Had}_{\mu\nu}(p,q)={\cal B}\Pi^{QCD}_{\mu\nu}(p,q),
\end{eqnarray}
which leads to
\begin{eqnarray}
{\cal B}\Pi_1^{Had}={\cal B}\Pi_1^{QCD},~~~~{\cal B}\Pi_2^{Had}={\cal B}\Pi_2^{QCD},
\end{eqnarray}
 corresponding to the structures  $g_{\mu\nu}\pslash\eslash\qslash$ and  $g_{\mu\nu}\eslash\qslash$. In this manner we extract the QCD sum rules for the form factors 
$F_1$ and  $F_2$. 
They are very lengthy functions, therefore we do not give their explicit expressions here.
The interested readers can find details of the calculations such as Borel transformations
and continuum subtraction in Refs.~\cite{Agaev:2016srl,Azizi:2018duk}

\section{Numerical analysis} \label{secIII}
In this section, we achieve numerical computations for 
the spin-$\frac{3}{2}$ DHBs. We use $m_u$ = $m_d$ = 0,  $m_s$ = $96^{+8}_{-4}$~MeV,  
$m_c$ =  $1.67 \pm 0.07$\,GeV, 
$m_b$ = $4.78 \pm 0.06$\,GeV, 
$f_{3\gamma}$ = $-0.0039~\text{GeV}^2$~\cite{Ball:2002ps},
$\langle \bar qq\rangle$ = $(-0.24\pm0.01)^3\,\text{GeV}^3$ \cite{Ioffe:2005ym},
$m_0^{2}$ =  $0.8 \pm 0.1~\text{GeV}^2$, $\langle g_s^2G^2\rangle = 0.88~ GeV^4$ 
 and $\chi$ = $-2.85 \pm 0.5~\text{GeV}^{-2}$~\cite{Rohrwild:2007yt}. 
The masses of the  $\Xi^{*}_{QQ}$, $\Xi^{*}_{QQ^{\prime}}$, $\Omega^{*}_{QQ}$ and $\Omega^{*}_{QQ^{\prime}}$
 baryons are borrowed from Ref.~\cite{Aliev:2012iv}, 
in which the mass sum rules have been used to compute them. 
These masses are used to have the following values: $M_{\Xi^{*}_{cc}}=3.72 \pm 0.18~\text{GeV}$, 
$M_{\Omega^{*}_{cc}}=3.78 \pm 0.16~\text{GeV}$, 
$M_{\Xi^{*}_{bc}}= 7.25 \pm 0.20~\text{GeV}$, 
$M_{\Omega^{*}_{bc}}= 7.30 \pm 0.20~\text{GeV}$, 
$M_{\Xi^{*}_{bb}}=10.40 \pm 1.00~\text{GeV}$ 
 and  $M_{\Omega^{*}_{bb}}=10.50 \pm 0.20~\text{GeV}$.
In order to specify the MDMs of DHBs, the value of
the residues are needed. The residues of the DHBs are computed in Ref.~\cite{Aliev:2012iv}.
These residues are calculated to have the following values: 
$\lambda_{\Xi^*_{cc}}=0.12 \pm 0.01~\text{GeV}^3$, 
$\lambda_{\Omega^*_{cc}}=0.14 \pm 0.02~\text{GeV}^3$, 
$\lambda_{\Xi^*_{bc}}=0.15 \pm 0.01~\text{GeV}^3$, 
$\lambda_{\Omega^*_{bc}}=0.18 \pm 0.02~\text{GeV}^3$, 
 $\lambda_{\Xi^*_{bb}}=0.22 \pm 0.03~\text{GeV}^3$ 
 and $\lambda_{\Omega^*_{bb}}=0.25 \pm 0.03~\text{GeV}^3$.
The parameters used in the photon distribution amplitudes are given in Ref. \cite{Ball:2002ps}.

\begin{figure}[t]
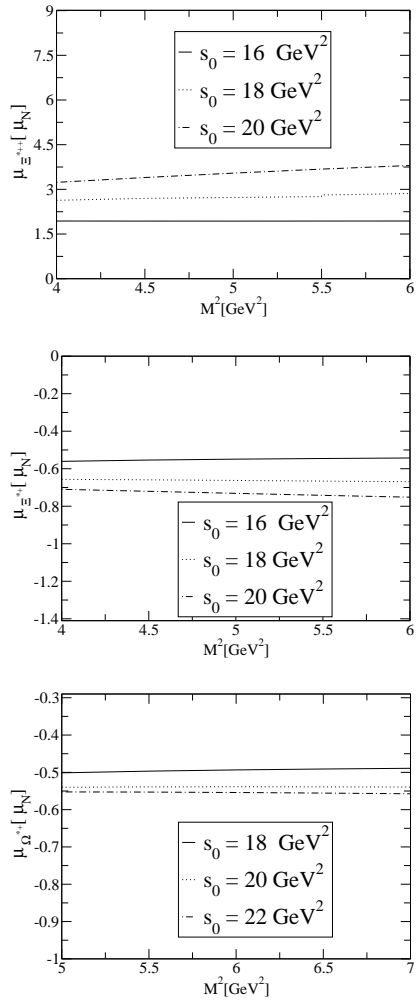

\centering
 \includegraphics[width=0.3\textwidth]{1XccuS-three.eps}\\
  \vspace{0.4cm}
 \includegraphics[width=0.3\textwidth]{1XccdS-three.eps}\\
    \vspace{0.4cm}
\includegraphics[width=0.3\textwidth]{1OccsS-three.eps}
 \caption{ 
 The dependence of the MDMs for the spin-$\frac{3}{2}$ doubly charmed baryons on the $M^{2}$ at various fixed values of the $s_0$.}
 \label{fig:Msq}
  \end{figure}

The QCD sum rule for the MDMs of the DHBs, besides the above mentioned input
parameters, include also two more extra parameters. These parameters are the continuum threshold $s_0$ and the Borel mass
parameter $M^2$. According to the QCD sum rules philosophy we
need to find the working regions of these parameters, where the MDMs of the DHBs be insensitive
to the variation of these parameters in their working regions.
The $s_0$ is not  entirely optional parameter, it is preferred as the point at which the  continuum and excited states begin to contribute to the calculations.
To decide the working interval of the $s_0$, we enforce the conditions of operator product expansion (OPE) convergence and pole dominance. 
Therefore, it is expected that $s_0$ varies in the interval $(M_{B^*_{QQ}}+0.3)^2 $ GeV$^2 \leq s_0 \leq (M_{B^*_{QQ}}+0.7)^2$ GeV$^2$.
From this point of view, we prefer the value of the $s_0$ within the interval
 $s_0 = (16-20)~\text{GeV}^2$ for $\Xi^*_{cc}$, 
  $s_0 = (58-62)~\text{GeV}^2$ for $\Xi^*_{bc}$, 
 $s_0 = (116-120)~\text{GeV}^2$ for $\Xi^*_{bb}$, 
 $s_0 = (18-22)~\text{GeV}^2$ for $\Omega^*_{cc}$, 
   $s_0 = (60-64)~\text{GeV}^2$ for $\Omega^*_{bc}$ and 
 $s_0 = (118-122)~\text{GeV}^2$ for $\Omega^*_{bb}$ baryons.
The working region for $M^2$ is achieved by requiring that the series of OPE in QCD representation is convergent and the contribution
of higher states and continuum is efficiently suppressed.
In technique language, the upper limit on $M^2$ is found demanding the maximum pole contribution. 
The lower limit is obtained demanding that the contribution of the 
perturbative part exceeds 
the nonperturbative one and  series of the operator product expansion in the obtained sum rules converge.
Our numerical calculations indicates that both conditions are satisfied
when $M^2$ changes in the regions: 
$4~\text{GeV}^2 \leq M^2 \leq 6~\text{GeV}^2$ for $\Xi^*_{cc}$,
$7~\text{GeV}^2 \leq M^2 \leq 9~\text{GeV}^2$ for $\Xi^*_{bc}$, 
$10~\text{GeV}^2 \leq M^2 \leq 14~\text{GeV}^2$ for $\Xi^*_{bb}$, 
$5~\text{GeV}^2 \leq M^2 \leq 7~\text{GeV}^2$ for $\Omega^*_{cc}$,
$8~\text{GeV}^2 \leq M^2 \leq 10~\text{GeV}^2$ for $\Omega^*_{bc}$  and 
$11~\text{GeV}^2 \leq M^2 \leq 15~\text{GeV}^2$ for $\Omega^*_{bb}$ baryons. 
As an example in Fig. \ref{fig:Msq}, we present the dependencies of the MDMs of doubly charmed baryons on $M^2$ at several fixed values of the $s_0$.  
As is seen from the figure, although being not entirely insensitive, the MDMs exhibit acceptable dependency on 
the extra parameters, $s_0$ and $M^2$  which is reasonable in the error limits of the QCD sum rule formalism.

Our final results on the MDMs for the spin-$\frac{3}{2}$ DHBs are
\begin{eqnarray}
\mu_{\Xi_{cc}^{*++}} &=& 2.94 \pm 0.95,\nonumber\\
\mu_{\Xi_{bc}^{*+}}  &=& 2.63 \pm 0.82, \nonumber \\
\mu_{\Xi_{bb}^{*0}}  &=& 2.30 \pm 0.55, \nonumber \\
\mu_{\Xi_{cc}^{*+}}   &=& - 0.67 \pm 0.11,\nonumber\\
\mu_{\Xi_{bc}^{*0}}   &=& - 0.96 \pm 0.32,\nonumber\\
\mu_{\Xi_{bb}^{*-}}   &=& -1.39 \pm 0.32,\nonumber\\
\mu_{\Omega_{cc}^{*+}} &=&- 0.52 \pm 0.07, \nonumber\\
\mu_{\Omega_{bc}^{*+}} &=&- 1.11 \pm 0.33,\nonumber\\
\mu_{\Omega_{bb}^{*-}} &=& -1.56 \pm 0.33,
\end{eqnarray}
where the quoted errors in the results are in connection with the uncertainties in the values of the input parameters and the photon distribution amplitudes, 
as well as the variations in the computations of the working windows $ M^2 $ and $ s_0 $.
We also need to emphasize that the main source of uncertainties is the variations with respect to $s_0$ and the results weakly depend on the
choices of the $M^2$.

%
In Table \ref{Table:comp}, we present the our numerical results for the MDMs and comparison of the with various other models 
such as,  
the relativistic harmonic confinement
model (RHM)~\cite{Gadaria:2016omw}, 
MIT bag model~\cite{Bernotas:2012nz,Simonis:2018rld}, 
nonrelativistic quark model (NRQM)~\cite{Albertus:2006ya}, 
hyper central constituent model (HCQM)~\cite{ Shah:2016vmd,Shah:2017liu}, effective mass (EMS) and screened charge
scheme (SCS)~\cite{Dhir:2009ax,Dhir:2013nka} and  
 heavy baryon chiral perturbation theory (HBChBT)~\cite{Meng:2017dni}.
From a comparison of our values with the predictions of other models we observe from this table that for the the $\Xi_{cc}^{*++}$  baryon,
practically all methods give, approximately, similar predictions. 
 For the $\Xi_{cc}^{*+}$ and $\Omega^{*+}_{cc}$ baryons, there are large discrepancy among results not only the magnitude but also by the sign.
 The reason for such inconsistencies is relatively easy to understand.
 The sign of the MDM depends on what is stronger -two heavy quarks and one light quarks. 
 In our analysis, the light quark overcome two heavy quarks and give the dominant contributions. 
For the the $\Xi_{bb}^{*0}$  baryon, our estimation is consistent within the errors with Refs.~\cite{Albertus:2006ya,Gadaria:2016omw,Meng:2017dni} and unlike other approaches. 
For the the $\Xi_{bb}^{*-}$  baryon, nearly all models give, approximately, similar predictions except the values of Refs.~\cite{Bernotas:2012nz,Simonis:2018rld}, which are small.
For the $\Omega^{*-}_{bb}$ baryon, our estimation is consistent within the errors with Refs.~\cite{Shah:2016vmd,Shah:2017liu,Meng:2017dni} and different from other results. 
For the the $\Xi_{bc}^{*+}$  baryon, we see that within errors our predictions in good agreement with the Refs~\cite{Albertus:2006ya,Gadaria:2016omw,Dhir:2009ax,Dhir:2013nka,Meng:2017dni}.
For the $\Xi^{*0}_{bc}$ baryon, while the sign of the MDM is correctly
determined, there is a large discrepancy among results.
For the $\Omega^{*0}_{bc}$ baryon, nearly all methods give, moderately, similar approximations except the results of Ref.~\cite{Meng:2017dni} and this work, which are quite large.
 \begin{widetext}
 
  \begin{table}[t]
\centering
 \caption{ MDMs of the spin-$\frac{3}{2}$ DHBs (in nuclear magnetons $\mu_{N}$).} 
    \label{Table:comp}
\begin{tabular}{l|c|c|c|c|c|c|c|c|c}
\hline \hline
 Approaches&~$\Xi_{cc}^{*++}$~&~$\Xi_{cc}^{*+}$~&~$\Omega_{cc}^{*+}$~&~$\Xi_{bb}^{*0}$~&~$\Xi_{bb}^{*-}$~&~$\Omega_{bb}^{*-}$~&~$\Xi_{bc}^{*+}$~&~$\Xi_{bc}^{*0}$~&~$\Omega_{bc}^{*0}$\\
 \hline \hline
NRQM~ 
\cite{Albertus:2006ya}                          & 2.67 & -0.31  & 0.14 & 1.87 &-1.11 &-0.66 &2.27&-0.71&-0.26\\
HCQM~
\cite{Shah:2016vmd,Shah:2017liu}                & 2.22 &  0.07  &0.29  &1.61  &-1.74 &-1.24 &1.56&-0.38&-0.18\\
HBChBT-I~
\cite{Meng:2017dni}                             & 3.51 & -0.27  &-0.64 &2.83  &-1.33 &-1.54 &3.22&-0.84&-1.09\\
HBChBT-II~
\cite{Meng:2017dni}                             & 3.63 & -0.37  &-0.65 &2.87  &-1.38 &-1.55 &3.27&-0.89&-1.10\\
EMS~
\cite{Dhir:2009ax,Dhir:2013nka}                 & 2.41 & -0.11  & 0.16 & 1.60 &-0.98 &-0.70 &2.01&-0.55&-0.28\\
SCS~
\cite{Dhir:2009ax,Dhir:2013nka}                 & 2.52 &  0.04  & 0.21 & 1.50 &-1.02 &-0.80 &2.02&-0.50&-0.30\\
MIT Bag model-I~
\cite{Bernotas:2012nz}                          & 2.00 &  0.16  & 0.33 &0.92  &-0.65 &-0.52 &1.41&-0.25&-0.11\\
MIT Bag model-II~
\cite{Simonis:2018rld}                          & 2.35 & -0.18  &-0.05 & 1.40 &-0.88 &-0.70 &1.88&-0.54&-0.33 \\
RHM~
\cite{Gadaria:2016omw}                          & 2.72 & -0.23  &0.16  &2.30  &-1.32 &-0.86 &2.68&-0.76&-0.32\\
This work                                       & 2.94 & -0.67  &-0.52 &2.30  &-1.39 &-1.56 &2.63&-0.96&-1.11\\
\hline \hline
\end{tabular}
  \end{table}
  
  \end{widetext}
 
 There is no experimental data for the MDMs of the DHBs. 
 But, one can define some useful splittings rules for the checking and comparing the results.
 In the heavy quark limit, $\Xi_{QQq}^{*}$, $\Xi_{QQ'q}^{*}$, $\Omega_{QQq}^{*}$ and $\Omega_{QQ'q}^{*}$ baryons have the same MDMs. 
Because they have the same light degrees of freedom and the contribution of heavy quarks will vanished in this limit. 
If one keep the heavy quark mass, the MDMs are not equal anymore. 
However one can expect that they may satisfy following relations as the $\Xi_{QQ'q}^{*}$ and $\Omega_{QQ'q}^{*}$ are intermediate states: 
\begin{eqnarray}
&& \Delta\mu_{u}\equiv\mu_{\Xi_{bbu}^{*0}}+\mu_{\Xi_{ccu}^{*++}}-2\,\mu_{\Xi_{bcu}^{*+}}= 0,\nonumber\\
&& \Delta\mu_{d}\equiv\mu_{\Xi_{bbd}^{*-}}+\mu_{\Xi_{ccd}^{*+}}-2\,\mu_{\Xi_{bcd}^{*0}}= 0,\\
&& \Delta\mu_{s}\equiv\mu_{\Omega_{bbs}^{*-}}+\mu_{\Omega_{ccs}^{*+}}-2\, \mu_{\Omega_{bcs}^{*0}}= 0.\nonumber
\end{eqnarray}

We observe from Table \ref{Table:comp1} that all baryons almost satisfy that condition except the results are obtained in Refs. \cite{Shah:2016vmd,Shah:2017liu}.
 \begin{table}[htp]
\centering
 \caption{ Comparision of the first splitting.}
\label{Table:comp1}
	\begin{tabular}{l|c|c|c}
	\hline \hline
	Approaches&~ $
	 \Delta\mu_{u}$ & $ \Delta\mu_{d}$ & $ \Delta\mu_{s}$  \\
	 \hline
	NRQM~\cite{Albertus:2006ya}                            & 0.00 & 0.00& 0.00            \\
	HCQM~\cite{Shah:2016vmd,Shah:2017liu}                  &0.71& -0.91& -0.59  \\
	HBChBT-I~\cite{Meng:2017dni}                           &-0.10& 0.08& 0.00 \\
	HBChBT-II~\cite{Meng:2017dni}                          &-0.04& 0.03& 0.00 \\
	EMS~\cite{Dhir:2009ax,Dhir:2013nka}         & -0.01& 0.01& 0.02 \\
	SCS~\cite{Dhir:2009ax,Dhir:2013nka}        &-0.02& 0.02& 0.01 \\
	MIT Bag model-I~\cite{Bernotas:2012nz}                 &0.10& 0.01& 0.03 \\
	MIT Bag model-II~\cite{Simonis:2018rld}                &-0.01& 0.02& -0.09 \\
	RHM~\cite{Gadaria:2016omw}                             &-0.34& -0.03& -0.06\\
	This work                                              &-0.02& -0.14&0.14\\
	\hline \hline
\end{tabular}
\end{table}

As we mentioned beginning of this section, we work on $m_{u}=m_{d}=0$ limit. 
The second splittings can be defined 
\begin{eqnarray}
&&\Delta{\mu}^{QQ}_{u-d}\equiv \mu_{\Xi^*_{QQu}}-\mu_{\Xi^*_{QQd}},\nonumber\\ 
&&\Delta{\mu}^{QQ'}_{u-d}\equiv \mu_{\Xi^*_{QQ'u}}-\mu_{\Xi^*_{QQd}},
\end{eqnarray}
which gives the difference occur as a result of the different charge of $u$ and $d$ quark. The third splittings  can also be defined,
\begin{eqnarray}
&&\Delta{\mu}^{QQ}_{d-s}\equiv \mu_{\Xi^*_{QQd}} - \mu_{\Omega^*_{QQs}},\nonumber\\ 
&&\Delta{\mu}^{QQ'}_{d-s}\equiv \mu_{\Xi^*_{QQ'd}} - \mu_{\Omega^*_{QQ's}},
\end{eqnarray}
 which gives the difference occur as a result of the different quark mass of $d$ and $s$.
In the second and third splittings, the heavy quark contribution has been canceled out. Therefore, $\Delta\mu_{u-d/d-s}^{cc}= \Delta\mu_{u-d/d-s}^{bb}= \Delta\mu_{u-d/d-s}^{bc}$ is expected. 
We can see most results in the Table \ref{Table:comp2} satisfy this relation, except the $\Delta\mu_{d-s}^{cc}$ of this work. 

\begin{widetext}
 
 \begin{table}[htp]
\centering
\caption{ Comparision of the second and third splittings.}
\label{Table:comp2}
	\begin{tabular}{l|c|c|c|c|c|c}
	\hline \hline
	Approaches&~ $ \Delta\mu_{u-d}^{cc}$ & $ \Delta\mu_{u-d}^{bb}$ & $ \Delta\mu_{u-d}^{bc}$&  $ \Delta\mu_{d-s}^{cc}$ & $ \Delta\mu_{d-s}^{bb}$ & $ \Delta\mu_{d-s}^{bc}$  \\
	\hline
	NRQM~\cite{Albertus:2006ya}                     &2.98& 2.98& 2.98& -0.45& -0.45& -0.45           \\
	HCQM~~\cite{Shah:2016vmd,Shah:2017liu}          &2.15& 3.35& 1.94& -0.22& -0.50& -0.20  \\
	HBChBT-I~\cite{Meng:2017dni}                    & 3.78& 4.16& 4.06& 0.37& 0.21& 0.25 \\
	HBChBT-II~~\cite{Meng:2017dni}                  & 4.00& 4.25& 4.16& 0.28& 0.17& 0.21 \\
	EMS~\cite{Dhir:2009ax,Dhir:2013nka}             & 2.52& 2.58& 2.56& -0.27& -0.28& -0.27 \\
	SCS~\cite{Dhir:2009ax,Dhir:2013nka}             &2.48& 2.52& 2.52& -0.17& -0.22& -0.20 \\
	MIT Bag model-I~\cite{Bernotas:2012nz}          &1.84& 1.57& 1.66& -0.17& -0.13& -0.14 \\
	MIT Bag model-II~\cite{Simonis:2018rld}         &2.53& 2.28& 2.42& -0.13& -0.18& -0.21 \\
	RHM~\cite{Gadaria:2016omw}                      & 2.95& 3.62& 3.44& -0.39& -0.46& -0.44\\
	This work                                       & 3.61& 3.69& 3.59& -0.15& 0.17& 0.15\\
	\hline \hline
\end{tabular}
 \end{table}

 \end{widetext}
 
 \section{Discussion and concluding remarks}\label{secIV}

In the presented paper we have evaluated the MDMs
 of the spin-$\frac{3}{2}$ DHBs by means of the light-cone QCD sum rule. 
The MDMs of the DHBs encodes 
key knowledge of their internal structure and shape deformations.
Measurement of the MDMs of the spin-$\frac{3}{2}$ DHBs
in future experiments can be very helpful understanding the internal structure of these baryons.
However, the direct measurement of the MDMs of the spin-$\frac{3}{2}$ DHBs are unlikely in the
near future. Therefore, any unstraightforward approximation of the MDMs of the spin-$\frac{3}{2}$ DHBs could be very helpful.
%
%
Comparison of our results with the estimation of other theoretical models is presented.
As can be seen from the MDM results of the DHBs given in Table \ref{Table:comp}, 
the results obtained using different models lead to rather different estimations,
which can be used to distinguish these models. 
Obviously, more studies are needed to understand the current situation.

\section{Acknowledgements}
We are grateful to V. S. Zamiralov for useful discussions, comments and remarks.


\bibliography{refs}

\end{document}